\documentstyle[11pt,newpasp,twoside,epsf]{article}

\begin{document}

\title{Binary Evolution and Neutron Stars in Globular Clusters}
\author{Natalia Ivanova, John M.\ Fregeau, and Frederic A.\ Rasio}
\affil{Department of Physics and Astronomy, Northwestern University,
       2131 Tech Drive, Evanston, IL 60208}

\begin{abstract}
We investigate the dynamical formation and evolution of binaries containing 
neutron stars in dense globular clusters. Our numerical simulations combine
a simple Monte Carlo prescription for stellar dynamics,
a sophisticated binary population synthesis code,
and a small-$N$-body integrator for computing 3-body and 4-body interactions. 
Our results suggest that there is no ``retention 
problem,'' i.e., that, under standard assumptions, globular clusters can 
retain enough neutron stars to 
produce the observed numbers of millisecond pulsars.
We also identify the dominant evolutionary and dynamical channels through which 
globular clusters produce their two main
types of binary millisecond pulsars.
\end{abstract}

\section{Introduction}

Globular clusters (GCs) have proven to be a gold mine for studies of
radio millisecond pulsars (MSPs), both isolated and in binaries:
observations show that the number of MSPs formed per unit mass in GCs 
is much higher than in the galactic field. 
The companion-mass and orbital-period distributions of the observed binary MSPs 
in GCs and in the field are rather different (see Fig.~1).
Clearly, the mechanisms responsible for MSP
formation in GCs and in the field must also be different.
Indeed, from the earliest X-ray observations, it was suggested 
that compact binaries in GCs must be formed through dynamical 
interactions in the high-density environment (Clark 1975).

Close stellar encounters can lead mainly to:
(i) the destruction of wide binaries; 
(ii) hardening of close binaries (following ``Heggie's law''; Heggie 1975);
(iii) exchange interactions, through which low-mass companions tend to be 
replaced by a more massive participant in the encounter;
(iv) physical collisions and mergers in binary--single or binary--binary 
encounters.
In addition, both single stars and binaries can be ejected from the cluster
through recoil following an encounter or 
supernova (SN) explosion.
All these processes naturally introduce a set of key theoretical questions:
(i) how many neutron stars (NSs) will remain in the cluster, 
(ii) how many of them can be recycled, 
(iii) how many can obtain or retain a binary companion and 
(iv) what are the expected characteristics of NS binaries?
Answers to these questions rely strongly on our understanding
of the binary stellar evolution coupled with cluster dynamics.

\begin{figure}
\plotfiddle{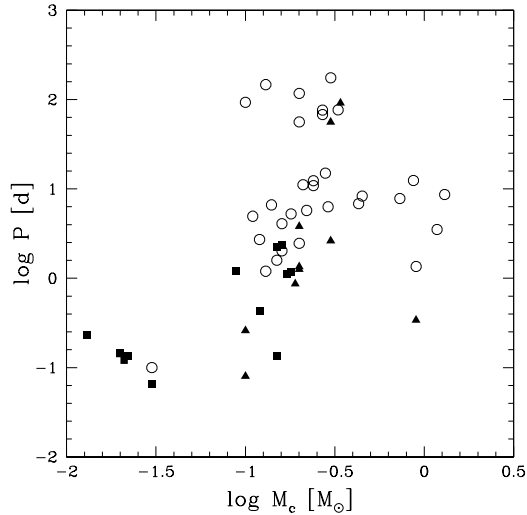}{6cm}{0}{35}{35}{-100}{-60}
\caption{Orbital period vs companion mass for all observed binary MSPs 
in the galactic field (open circles), in 47~Tuc (solid squares) and in
other globular clusters (solid triangles).} 
\end{figure}

%
In our current work we use a new {\it Monte Carlo\/} approach (Sec.~2)
based on one of the most advanced  
binary population synthesis codes currently available ({\tt StarTrack}; see Belczynski, Kalogera \& Bulik 2002).
The cluster is modeled as a fixed background:
all relevant parameters such as central density and velocity dispersion, and
half-mass relaxation time (but not total mass),
are kept constant throughout each dynamical simulation.
This assumption is well justified physically for GCs with significant fractions of primordial binaries,
which can be fitted by standard King models (see, e.g., Fregeau et al.\ 2003) 
and has been
used in many previous theoretical studies of binary interactions
and cluster dynamics (e.g., Hut, McMillan \& Romani 1992).
The most important dynamical processes treated by the code include mass
segregation and evaporation, physical collisions, tidal captures, and binary--single
and binary--binary encounters.
Each dynamical encounter involving a binary is calculated using
{\tt Fewbody}, a new numerical toolkit for direct $N$-body integrations 
of small-$N$ gravitational
dynamics that is particularly suited to performing 3-body and 4-body
integrations with high accuracy (Sec.~6).

Our results on the retention of NSs and on the statistics of NS populations 
in GCs are described in Sec.~3.
In Sec.~4 we discuss the main types of NS binaries that are
formed dynamically, and their survival probability and further evolution 
in the cluster. We then
compare these results to the observed MSP cluster population. 

\section{Method}

We start our  typical simulations with $N\sim 10^5$ stars and with
the initial binary fraction in the range 50\% -- 100\%. 
This high primordial binary fraction
(much higher than assumed in all previous studies) 
is needed in order to match the observed binary 
fractions in GC cores today (Ivanova et al.\ 2004).
Here we consider two representative core densities: 
$\rho_{\rm c} =10^{4.5}\, M_\odot\,{\rm pc}^{-3}$ (``typical'' cluster model) and
$\rho_{\rm c} =10^{5.1}\, M_\odot\,{\rm pc}^{-3}$ (``47~Tuc'' cluster model).
The one-dimensional velocity dispersion in the core is assumed to be $\sigma_1=10\,$km/s 
and the corresponding escape speed from the cluster $v_{\rm esc} = 60\,$km/s.

Following  Rasio, Phahl \& Rappaport (2000),
we assume that the probability for an object of mass $m$ 
to enter the cluster core after a time $t_{\rm s}$ follows a Poisson distribution,
$p(t_{\rm s})=(1/t_{\rm sc})\exp(-t_{\rm s}/t_{\rm sc})$, where the characteristic
mass-segregation timescale is given by $t_{\rm sc}=10 \left(\langle m
\rangle/m\right) t_{\rm rh}$  (Fregeau et al.\ 2002). 
The half-mass relaxation time is taken to be constant for a given cluster
and for all simulations here we set $t_{\rm rh}= 1\,$Gyr. 
We adopt the broken power-law IMF of Kroupa (2002) for single stars
and primaries, a flat binary-mass-ratio distribution for secondaries, and
the distribution of initial binary periods constant in logarithm
between contact and $10^7\,$d.

To evolve single stars, we use the analytic fits 
of Hurley, Pols, \& Tout (2000).
The binary evolution is calculated employing the 
binary  population synthesis code {\tt StarTrack}, which has been
used previously in many studies of double compact objects 
(e.g., Belczynski et al.\ 2002) and X-ray binaries
(e.g., Belczynski et al.\ 2004). 
This code allows us to follow the evolution of
binaries with a large range of stellar masses and metallicities.
It incorporates detailed treatments of stable or unstable
conservative or non-conservative, mass transfer (MT) episodes, mass and angular
momentum loss through stellar winds (dependent on metallicity) and
gravitational radiation, asymmetric core collapse events with a realistic
spectrum of compact object masses, and the effects of magnetic braking and
tidal circularization. 

The evolution of single and binary stars can be altered drastically
by dynamical
encounters with other objects. In our simulations, we consider such outcomes
of dynamical encounters as physical collisions, tidal captures, 
destruction of binaries, companion exchanges, triple formation, etc. 
A particularly important type of outcome that we take into account is 
the dynamical common envelope (CE) phase that follows a physical 
collision between a compact object and a red giant (RG). 
In particular, if this compact object is a NS, 
a binary containing the NS and a white dwarf (WD) companion can be 
formed (Rasio \& Shapiro 1990, 1991; Davies et al.\ 1991).
In this case, during the CE phase, the NS may accrete enough material from 
the envelope to become recycled as a MSP (Bethe \& Brown 1998).

\section{Numbers and Characteristics of Retained Neutron Stars}

We adopt the double-peaked distribution of natal NS kick velocities 
from Arzoumanian, Chernoff, \& Cordes (2002). 
With this choice we find that 6\% of all NSs are retained in our typical 
cluster simulation, i.e., 240 NSs in a cluster with a current total mass of 200,000$\,M_\odot$. For our 47~Tuc model 
(with $\sim 1.5\times 10^6\,M_\odot$), the number of retained NSs is 1750.
Moreover, if NSs can be formed through accretion induced collapse (AIC) of 
WDs, the fraction of NSs remaining in the GC can be as high as $\sim 10\%$.

Throughout most of the cluster evolution, retained NSs (single or in binaries)
are more massive than other objects. Their interaction cross section is large and 
they actively participate in dynamical encounters.
For example, the probability for any single NS to have a physical collision with a 
$\sim 1\,M_\odot$ MS star in our typical cluster model during 14 Gyr is about 10\%. Being in a binary
increases the probability of interaction.
As a result, 85\% of all retained NSs experienced some kind of strong 
interaction (binary destruction, acquiring a new companion,
significant binary orbit change, or physical collision) after the formation 
of the NS. Most of these interactions involve binaries
(60\% binary--single and 35\% binary--binary) and
5\% were single--single encounters (mainly physical collisions,
and a negligible number of tidal captures).
In about 20\% of binary encounters with a NS as a participant, 
a physical collision happened;
45\% of the NSs that interacted acquired a new binary companion; 
and 20\% of them acquired a new companion twice.
As a result, at the end of the simulation (cluster age of 14\,Gyr),
only $\sim 5\%$ of retained NS in binaries remain in 
their original binary system.
In our typical model, 25\% of all NSs remaining in the cluster 
at 14\,Gyr are in binaries 
(essentially all formed by dynamical encounters) 
while 75\% of NSs are single;
about $5\%$ of all NSs are MSPs and $40\%$ of MSPs are in binaries;
75\% of all MSPs were first recycled in the original primordial binary.
In our 47~Tuc model the fraction of recycled NSs is higher, about $13$\%, 
and the total number of MSPs is 230, with 90 
binary MSPs. These numbers are comparable
to the estimated numbers of MSPs in 47~Tuc derived from observations
(Heinke et al.\ 2003).
Thus we conclude that there is no ``retention problem'' in our model.

\section{Properties of Dynamically Formed Binaries}

As mentioned above, our results show that most NS-binaries present 
today in a globular cluster were formed dynamically.
The survival and fate of these binaries depends on the
characteristic time to the next encounter (the collision time $t_{\rm coll}$) 
and on the probability for the NS to remain in the binary after the next encounter.
The latter depends on the hardness of the binary and on the masses
of the participating stars. In particular, soft binaries (having
binding energy smaller than the average kinetic energy ``at infinity''
of an encounter) will be easily destroyed (``ionized''). 
In a hard binary, encounters can lead to further hardening and/or  
exchanges. The NS is often the most massive participant, in which case it
will likely remain in the binary). As the NS-binary hardens, 
the probability of a physical collision increases
(Fregeau et al.\ 2004), and destruction can then occur through mergers.

\begin{figure}
\plotone{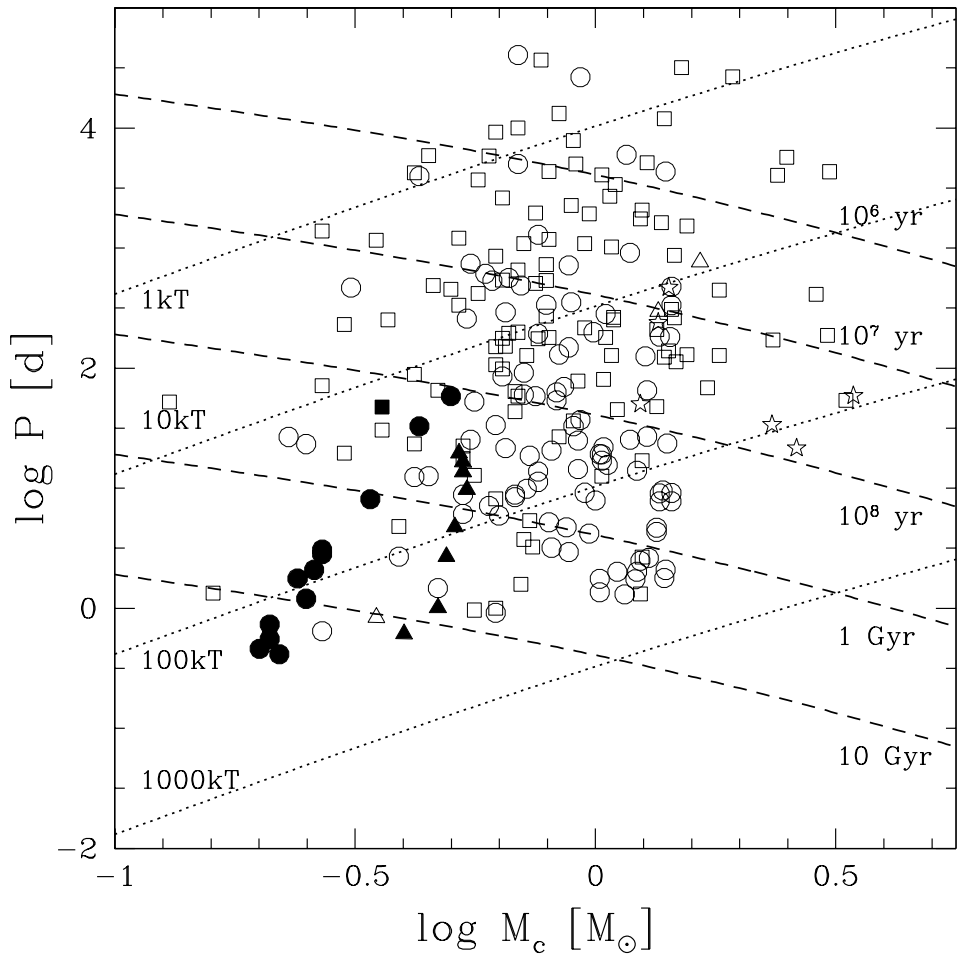}
\caption{Orbital periods vs companion masses for all
dynamically formed NS binaries during the entire evolution of our
47~Tuc model, normalized to a total mass of $3\times10^5\,M_\odot$. 
Values plotted are at the time of formation
(squares: MS companions; stars: RGs; triangles: He stars; circles: WDs).
Open symbols are for binaries formed by exchange interactions,
filled symbols for physical collisions.
Dashed lines have constant $t_{\rm coll}$ and dotted
lines have constant binary hardness (where ``1kT'' is a marginally hard
binary).}
\end{figure}

\begin{figure}
\plotfiddle{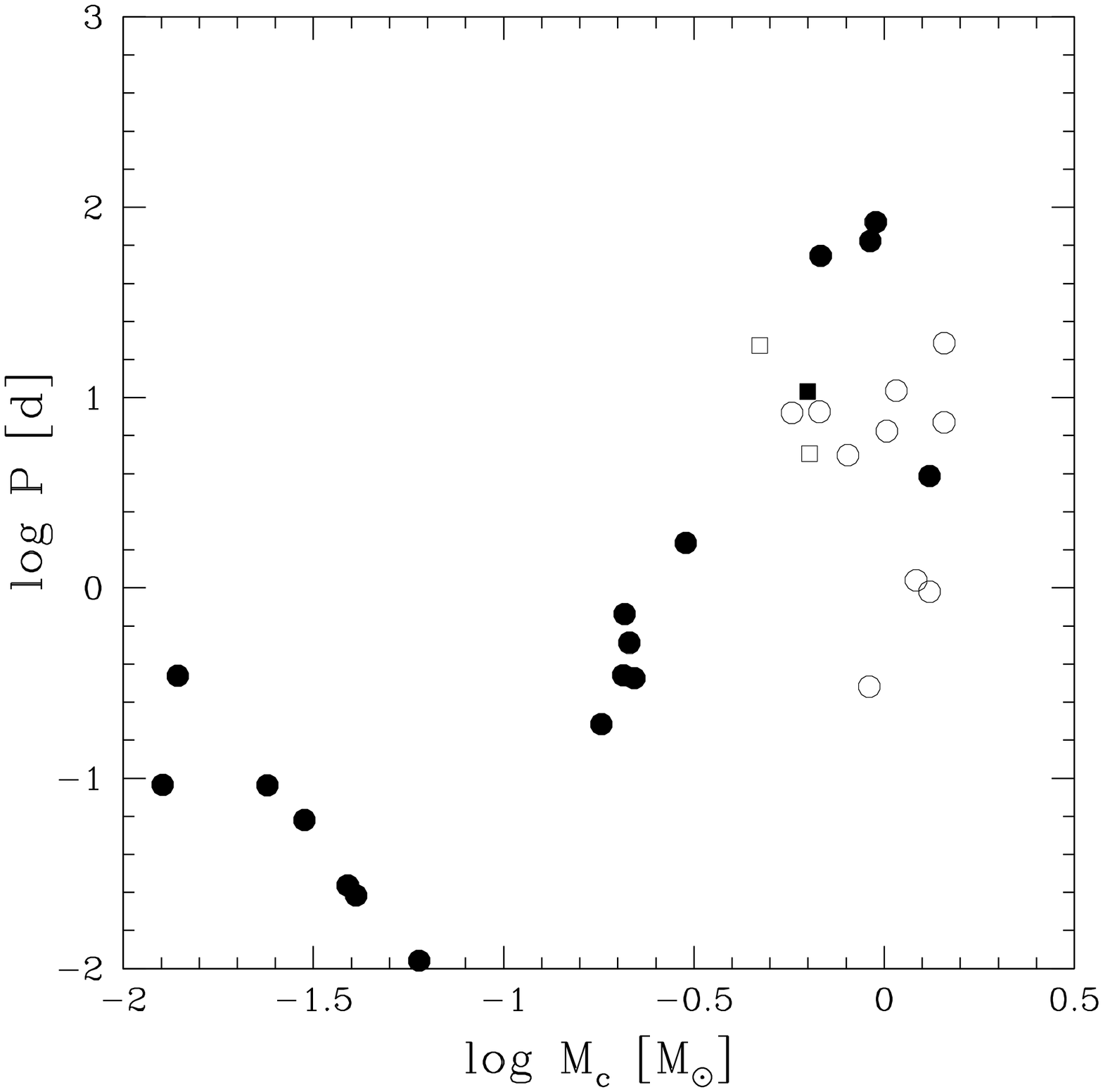}{6cm}{0}{35}{35}{-100}{-60}
\caption{Orbital periods vs companion masses for all
NS binaries in our 47~Tuc model at 14\,Gyr. Squares represent NSs in
binaries with MS companions, circles with WD companions.
Filled symbols represent binary MSPs. }
\end{figure}

In Figure~2 we show all binaries that were formed during the
entire evolution of our model with 47~Tuc-like parameters.
Most of the binaries were formed with a WD
or a MS star as a companion to the NS, following an exchange interaction. 
These binaries are typically formed with high
eccentricities and can shrink their orbit very fast 
through tidal dissipation and/or gravitational radiation.
Not all of these systems will survive after the onset of MT: if a 
WD-companion is more massive than about $0.6\,M_\odot$, the MT is unstable and 
the binary will merge. A significant fraction of the hardest
binaries (hardness ratio $\ga 100\,$kT) with very long collision times ($\ga 1\,$Gyr)
were formed through physical collisions of a NS and a RG. 
In these collisions a dynamical CE occurs and 
a close binary with a MSP and a WD or He-star companion is formed. 

In our typical cluster model, 
35\% of all NSs dynamically acquired a binary companion and about half of these
acquired a new companion at least twice (on average, every NS that formed a binary
had 3.5 exchange interactions). Most of the resulting binaries are hard, but
60\% of them are destroyed by a subsequent encounter, and
10\% are destroyed by evolutionary mergers (this fraction increases for denser
clusters, where, on average, harder binaries are formed).

As shown in Figure~2, most of the dynamically formed binaries have relatively short
collision times ($t_{\rm coll} \la 1\,$Gyr), while many of those with longer collision times
will merge once MT starts. As a result, very few of these
NS-binaries remain at 14\,Gyr (see Fig.~3).
Most of the remaining NS in binaries have a WD as a companion.
Binaries containing a recycled NS are divided into two main groups.
The first has very short periods and very small companion masses (lower-left corner of Fig.~3): these systems
evolved through stable MT, which followed an exchange interaction
of the NS into a WD--WD or WD--MS binary. The resulting NS--WD binary does not necessarily
have a very short period after the exchange. However, high eccentricities 
help shrink the binary orbits.
These NS--WD binaries also often experience subsequent encounters 
which lead to further hardening and eccentricity pumping, 
and possible further exchanges. 
The other group, with WD companions of $\sim 0.2\,M_\odot$ and longer 
periods in the range 0.1--0.3\,d, comes
mainly from physical collisions between a NS and a RG.
The resulting binary MSP distribution looks remarkably similar to the observed one
(see Fig.~1).

\section{Conclusions}

The formation and evolution of NS binaries in a dense stellar system 
almost always involves dynamical interactions. 
In a typical cluster, we find that only a few percent of NS binaries 
are of primordial origin. Our simulations also suggest that,
with a realistic NS kick velocity distribution, globular clusters 
can retain a sufficient number of NSs to explain the observed numbers 
of MSPs. For a cluster like 
47~Tuc, we predict $\sim200$ MSPs, with roughly half of them in binaries.
We can produce the two main types of observed binary MSPs: with $\sim0.2\,M_\odot$
WD companions, and with very low-mass companions 
($\sim 0.02\,M_\odot$) and ultra-short periods. The first type comes mainly from 
NS--RG collisions 
followed by CE evolution, the second from exchange interactions.

\section{Appendix: {\tt Fewbody}}

Small-$N$ gravitational dynamics, such as binary--binary and binary--single interactions,
are important for the formation and evolution of binaries in GCs,
as well as in driving the global cluster evolution (see, e.g., Heggie \& Hut 2003).
{\tt Fewbody} is a new, freely 
available\footnote{See {\tt http://www.mit.edu/\~{}fregeau}, or search the web for ``Fewbody''.} 
numerical toolkit for simulating small-$N$ gravitational dynamics 
(Fregeau et al.\ 2004).  
It is a general $N$-body dynamics code, although it was written 
for the purpose of performing scattering experiments, and therefore has several
features that make it well-suited for that purpose.  
{\tt Fewbody} may be used standalone to compute individual small-$N$ 
interactions, but is generally used as a library of routines from within
larger numerical codes that, e.g., calculate cross sections, or evolve GCs
via Monte Carlo methods.

{\tt Fewbody} uses an adaptive integrator, optionally with global
pairwise  Kusta\-anheimo-Stiefel regularization (Heggie 1974; Mikkola 1985),
to advance the particle positions with time.
It uses a binary-tree algorithm to classify the $N$-body system into a set of hierarchies,
and uses the approximate analytical criterion of Mardling \& Aarseth (2001) for the
dynamical stability of triples, applied at each level in the binary tree, to approximately assess the stability of each hierarchy. 
It also uses a binary-tree algorithm to speed up integration
of $N$-body systems with multiple timescales, by isolating weakly perturbed hierarchies
(typically hard binaries) from the integrator.  {\tt Fewbody} uses a set of simple rules to
automatically terminate calculations when the separately bound hierarchies comprising
the system will no longer interact with each other or evolve internally.  Finally, 
{\tt Fewbody} performs collisions between stars during the integration in the 
``sticky-star'' approximation, with the radius of a collision product parameterized
by a single expansion factor.

As a simple example of the use of {\tt Fewbody}, consider the survivability of a 
pulsar--black hole (BH)
binary in a GC core.  Although there is now ample evidence for the possible
existence of intermediate-mass black holes (IMBHs) in the cores of some GCs
(e.g., van der Marel et al.\ 2002; Gerssen et al.\ 2002, 2003), 
a stellar-mass BH has never been observed, directly or indirectly, in a 
Galactic GC.
It has been suggested that BH--PSR binaries should exist in GCs, and that they
may provide the most likely means of detection of a stellar-mass BH in a cluster 
(Sigurdsson 2003).  
An obvious question suggests itself:  What is the survivability of such binaries in clusters?  Due to their large
interaction cross section, binaries are likely to interact with other stars (or binaries if the binary fraction is
large enough).  A binary--single interaction can destroy the BH--PSR binary in several ways.  
For example, a main sequence (MS) field star can exchange into the binary, ejecting the NS and leaving a BH--MS binary; or 
the MS field star and NS can collide, perhaps yielding a short-lived
Thorne-\.Zytkow-like object (Thorne \& \.Zytkow 1977).

\begin{table}[t]\small
  \begin{center}
    \caption{\footnotesize Normalized destruction cross sections and characteristic lifetimes of
      BH--PSR binaries in globular clusters.
      $\widetilde\sigma_{\rm destr}$ is the normalized total cross section for destruction 
      (clean exchange and the merger of the MS and NS),
      $\widetilde\sigma_{\rm exch}$ is the cross section for clean
      exchange (an exchange in which there were no collisions),
      $\widetilde\sigma_{\rm NS-MS}$ is the cross section 
      for the collision of the NS and MS, and hence the formation of a Thorne-\.Zytkow-like object,
      and $\tau$ is the characteristic lifetime as a function of the number density $n_5 = n/10^5 pc^{-3}$.}
    \vskip20pt
    \begin{tabular}{cccccc}
      \tableline 
      $m_{\rm BH}$ ($M_\odot$)& $a$ (AU)& 
         $\widetilde\sigma_{\rm destr}/(\pi a^2)$ & 
         $\widetilde\sigma_{\rm exch}/(\pi a^2)$ & 
         $\widetilde\sigma_{\rm NS-MS}/(\pi a^2)$  & $\tau$ (yr)\\
      \tableline
      10 & 1   & $0.30\pm0.02$ & $0.25\pm0.02$ & $0.024\pm0.006$  & 
        $2.3\times 10^8 n_5^{-1}$\\
      3  & 0.1 & $1.16\pm0.08$ & $0.28\pm0.04$ & $0.72\pm0.06$  
	& $1.7\times 10^9 n_5^{-1}$\\
      \tableline 
    \end{tabular}
  \end{center}
\end{table}

For hard binaries in GCs, the binary interaction cross section is dominated
by gravitational focusing, and so we define the normalized cross section as
$\widetilde\sigma = \sigma(v_\infty/v_c)^2$,
where $\sigma$ is the (non-normalized) cross section, $v_\infty$ is the relative velocity 
between the binary and intruder at infinity, and $v_c$ 
is the critical velocity.  The critical velocity is defined so that the total energy of the
binary--single system is zero at $v_\infty = v_c$.  
Using {\tt Fewbody}, we have 
performed many binary--single scattering encounters between a BH--PSR binary and a MS
intruder star to calculate binary destruction cross sections.  We vary the mass of the BH and semimajor axis of the 
binary, but keep all other parameters fixed.  We set $m_{\rm NS}=1.4\,M_\odot$, $R_{\rm NS}=15\,{\rm km}$,
$m_{\rm MS}=1\,M_\odot$, $R_{\rm MS}=1\,R_\odot$, $e=0.9$, and $v_\infty=7\,{\rm km}/{\rm s}$.  Table~1
lists normalized cross sections for two different BH--PSR binaries: one containing a typical primordial
stellar-mass BH, with $m_{\rm BH}=10\,M_\odot$ and $a=1\,{\rm AU}$; and the other containing a lower-mass BH with $m_{\rm BH}=3\,M_\odot$ and 
$a=0.1\,{\rm AU}$.   It is clear that for the wider binary, with $a=1\,{\rm AU}$, the destruction cross section is 
dominated by clean exchange.  However, for the tighter binary, with $a=0.1\,{\rm AU}$, the destruction
cross section is dominated by NS--MS mergers.

Using the cross section, one may calculate the destruction rate per binary, and hence
the lifetime, as
$\tau = 1/R_{\rm destr} = 1/(n \sigma_{\rm destr} v_\infty)$,
where $n$ is the number density of single stars, $\sigma_{\rm destr}$ is the destruction cross section, 
and $v_\infty$ is the relative velocity between the binary and single star at infinity.
In a dense cluster core the lifetime can be relatively short, as shown in Table~1.
BH--PSR binaries with $m_{\rm BH}=10\,M_\odot$ 
are expected to form at the rate of $\sim 1$ per GC.
Adopting $n=10^4\,{\rm pc}^{-3}$ as a typical core number density implies that there are likely to be
$\sim 10$ observable BH--PSR binaries with primordial BHs in our Galactic GC
system. With $m_{\rm BH}=3\,M_\odot$, BH--PSR binaries 
are likely to form only in the densest cluster cores.  Assuming again that
$\sim 1$ is formed per dense cluster, we adopt the lifetime listed in Table~1, and 
now find $\sim 1$ observable BH--PSR binaries.

\acknowledgements
We thank the Aspen Center for Physics for financial support. This work was 
supported in part by a Chandra Theory grant, NASA ATP Grant NAG5-12044, and NSF Grant
AST-0206276.

\end{document}